\documentclass[sigconf]{acmart}
\usepackage{tabularx}
\usepackage{subcaption}

\AtBeginDocument{%
  \providecommand\BibTeX{{%
    \normalfont B\kern-0.5em{\scshape i\kern-0.25em b}\kern-0.8em\TeX}}}

\copyrightyear{2022}
\acmYear{2022}
\setcopyright{rightsretained}
\acmConference[CHI '22]{CHI Conference on Human Factors in Computing Systems}{April 29-May 5, 2022}{New Orleans, LA, USA}
\acmBooktitle{CHI Conference on Human Factors in Computing Systems (CHI '22), April 29-May 5, 2022, New Orleans, LA, USA}
\acmDOI{10.1145/3491102.3502042}
\acmISBN{978-1-4503-9157-3/22/04}

\begin{document}

\title{Learning to Denoise Raw Mobile UI Layouts for Improving Datasets at Scale}

\author{Gang Li}
\email{leebird@google.com}
\affiliation{%
  \institution{Google Research}
  \city{Mountain View}
  \country{United States}
}
\author{Gilles Baechler}
\email{baechler@google.com}
\affiliation{%
  \institution{Google Research}
  \city{Zurich}
  \country{Switzerland}
}

\author{Manuel Tragut}
\email{mtragut@google.com}
\affiliation{
  \institution{Google Research}
  \city{Zurich}
  \country{Switzerland}
}

\author{Yang Li}
\email{liyang@google.com}
\affiliation{
  \institution{Google Research}
  \city{Mountain View}
  \country{United States}
}

\renewcommand{\shortauthors}{Li et al.}

\begin{abstract}
The layout of a mobile screen is a critical data source for UI design research and semantic understanding of the screen. 
However, UI layouts in existing datasets are often noisy, have mismatches with their visual representation, or consists of generic or app-specific types that are difficult to analyze and model. 
In this paper, we propose the CLAY pipeline that uses a deep learning approach for denoising UI layouts, allowing us to automatically improve existing mobile UI layout datasets at scale.
Our pipeline takes both the screenshot and the raw UI layout, and annotates the raw layout by removing incorrect nodes and assigning a semantically meaningful type to each node. To experiment with our data-cleaning pipeline, we create the CLAY dataset of 59,555 human-annotated screen layouts, based on screenshots and raw layouts from Rico, a public mobile UI corpus. Our deep models achieve high accuracy with F1 scores of 82.7\% for detecting layout objects that do not have a valid visual representation and 85.9\% for recognizing object types, which significantly outperforms a heuristic baseline. Our work lays a foundation for creating large-scale high quality UI layout datasets for data-driven mobile UI research and reduces the need of manual labeling efforts that are prohibitively expensive. 
\end{abstract}



\maketitle

\begin{figure*}[ht]
  \centering
  \includegraphics[width=1.\linewidth]{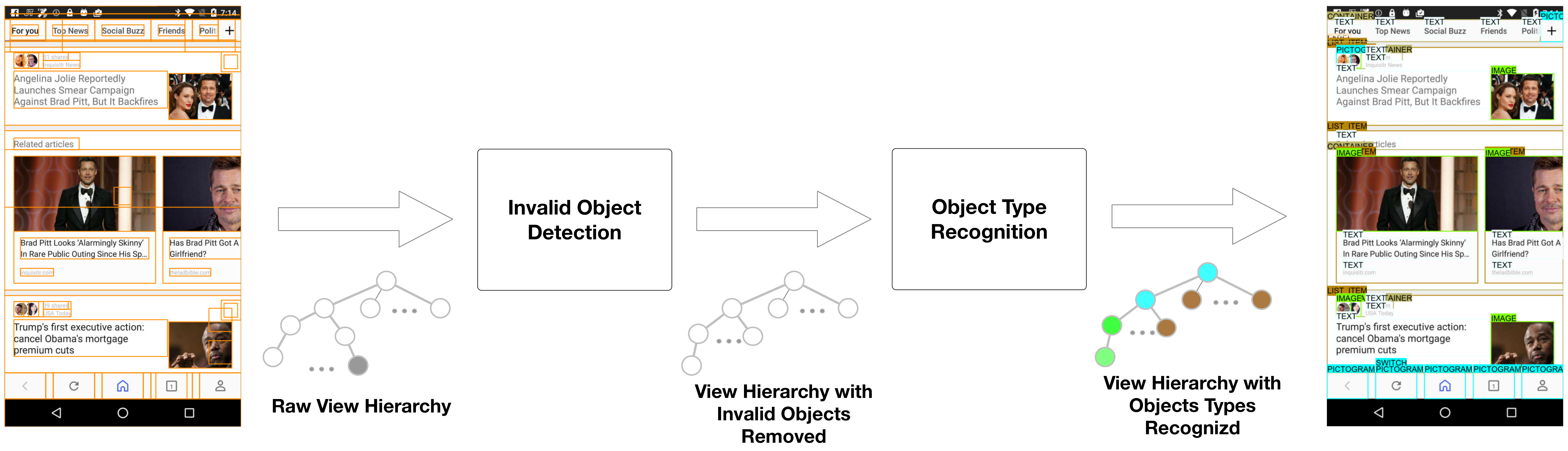}
  \caption{The CLAY pipeline cleans a mobile UI layout by first detecting objects in the layout structure (e.g., an Android View Hierarchy) that do not have a valid visual representation in the screenshot via 1) the Invalid Object Detection module, and then predicting the type of each object via 2) the Object Type Recognition module. The final output is a clean view hierarchy where each node matches the visual representations in the screenshot and is equipped with a well-defined object type.}~\label{fig:workflow}
  \Description{An overview of the proposed two-stage CLAY pipeline for denoising UI layouts. The pipeline first detects and removes invalid objects from the layout, and then assigns a more semantically meaningful type to the valid objects.}
\end{figure*}

\section{Introduction}\label{sec:introduction}

As mobile apps become prevalent in people's daily life, there has been a growing research interest in developing various machine learning applications based on mobile screens, e.g., UI component detection~\cite{Zhang2021-yo}, screen embedding~\cite{screen2vec}, widget captioning~\cite{Li2020-la}, icon annotation~\cite{zang2021multimodal} and screen summarization~\cite{Wang2021-bh}, which enhance the interactive capability and accessibility of mobile phones. These tasks usually rely on the screen layout\footnote{We use "layout" and "view hierarchy" interchangeably in this paper.}, i.e., the tree structure representation underlying the UI that contains the objects on the screen. Each object comes with a set of attributes such as its bounding box and type, which are often used either as input signal or output labels. For example, the widget captioning task~\cite{Li2020-la} takes in the cropped pixels of the objects based on their bounding boxes and their attributes including types from the view hierarchy, and outputs a brief caption describing the functionality of the object. 

However, UI layouts in existing datasets are often noisy. We have observed two major issues with these UI layouts. First, they are typically captured from the rendering tree or the accessibility tree of the UI at runtime. Similar to taking a screenshot, the results can be dynamic and out of sync. As a result, there can be invisible objects with bounding boxes with no visual correspondences on the screen,  misaligned objects with bounding boxes only partially covering the rendered objects, or objects in the background that are grayed out and not clickable. We refer to these objects in a layout as \emph{invalid} objects in the paper. In the Rico dataset~\cite{rico} we analyzed and annotated, about 37.4\% of the screens contain invalid objects. Second, the types of the objects in these captured view hierarchies can be either too generic, e.g., the \texttt{View} class in Android\footnote{\url{https://developer.android.com/reference/android/view/View}}, or too app-specific,  e.g., \texttt{ColombiaNativeAdView}, to be meaningful. Such types convey little information about the object for machine learning or data science tasks. In fact, the number of different types is virtually unlimited, which makes rule-based approaches hard to implement and extremely challenging to generalize. In the Rico dataset alone, we counted $9,331$ different types (view classes) in 59,555 view hierarchies. For comparison, the heuristic approach proposed in~\cite{Liu2018} only handles and maps $46$ such types to a semantic label.

The invalid objects and the problematic object types are not helpful as either input signal or output labels, and might even harm the model performance. For example, there has been recent interest in developing screen recognition/parser models \cite{Zhang2021-yo, screenparser} using the screenshot as input only. The objects in the view hierarchy can be potentially used as labels for these visual models. However, the invisible/misaligned objects are wrong labels to the visual models, as they don't match the rendered objects on the screen visually. Furthermore, the too generic object types (e.g., \texttt{View}) might include semantically different objects, which might confuse the visual models. The too specific object types would be too sparse for the models to learn and generalize. Similar problems exist when the objects are used as input features to other UI models (e.g., widget captioning \cite{Li2020-la}, language grounding \cite{li2020mapping}). The invisible/misaligned objects would result in invalid visual features and the too generic/specific object types would convey little or too sparse information for learning. Traditionally, these issues are addressed by employing human labelers to annotate a screen layout~\cite{Zhang2021-yo} to acquire clean layouts. However, the process is expensive because there are often tens of objects and containers on each screen to label. 

To address this issue, we develop the CLAY pipeline (see Figure~\ref{fig:workflow}) that employs deep learning models for automatically correcting a raw UI layout, by removing invalid objects and assigning a meaningful type to each object in the layout. Our pipeline consists of two steps: invalid object detection and object type recognition. Our invalid object detection module identifies and filters out the invalid objects. For object type recognition, we develop a multi-class classification model that is trained to assign to each object a meaningful type label, e.g., Drawer or Toolbar. To train and evaluate these models, we create a new dataset of 59,555 human-labeled screen layouts, based on the public Rico dataset~\cite{rico}. We design the invalid object detection model based on ResNet~\cite{he2015deep}. The multi-class classification models are based on two popular model architectures, namely Graphical Neural Networks (GNNs)~\cite{Gilmer2017} and Transformer~\cite{detr2020}, combined with a ResNet backbone. We demonstrate that the models achieved high accuracy and outperform a heuristic-based baseline on the test set, which show a great potential for denoising mobile UI datasets at scale. In summary, the paper makes the following contributions.

\begin{itemize}
    \item We identify the two main problems with existing mobile UI datasets: invalid objects and objects with generic or app-specific types. We propose a denoising task for mobile UI layouts, which can improve a dataset at scale.
    \item We create the large CLAY dataset of 59,555 screen layouts based on screenshots and raw layouts in Rico~\cite{rico}. Each object is either flagged as invalid, or labeled with a meaningful type. The dataset\footnote{The dataset and codes are released at https://github.com/google-research/google-research/tree/master/clay.} can be used as a common base for future model development and evaluation.
    \item We design a two-phase approach (the CLAY pipeline) for the denoising task. The first phase is a ResNet-based~\cite{he2015deep} filtering model, which performs binary classification for detecting invalid objects. The second phase is a multi-class classification model for the valid objects, based on Graphical Neural Networks and Transformers~\cite{detr2020}. We obtain 82.7\% and 85.9\% F1 scores for the first and second phase, respectively.
\end{itemize}

In the rest of the paper, we first discuss the literature related to our work in Section~\ref{sec:related_work}, and we then describe the dataset and the annotation process in Section
~\ref{sec:datasets}. The model architectures and experiment setup/results are presented in Section~\ref{sec:model_architecture} and~\ref{sec:experiments}, respectively. Finally, we discuss the limitations and propose directions for future work in Section~\ref{sec:discussions}.






\section{Related Work}\label{sec:related_work}

Our work is related to several research areas, including mobile UI dataset quality, UI screen modeling and data-driven automatic cleaning tools.

\subsection{Mobile UI Dataset Quality}
Datasets are the foundation for modeling and data science research. We are not the first ones to identify limitations in mobile UI layout datasets. 
Previously, Liu et al.~\cite{Liu2018} attempted to infer UI component types of elements from Android class names using a set of heuristics. However, they realized that objects with generic types make it difficult to produce reliable layout classes.
Li et al.~\cite{li2020mapping} found that there are view hierarchies out of sync with the screenshots and asked human raters to remove these erroneous screens; Li et al. \cite{Li2020-la} filtered out and discarded screens with inaccurate view hierarchy for further data labeling. Similar to Android UI datasets, there are similar issues with iOS datasets. \citet{Zhang2021-yo} noted that APIs for generating the screen layouts might have incomplete access to the UI metadata, and they asked human raters to manually label the layout of screens.

Recently, Fu et al.~\cite{fu2021understanding} recognized the need for cleaner layouts, and combined optical character recognition (OCR) with a graphic detector to recreate less cluttered layouts that are visually better synchronized with their screenshot. Another direction is to use object detection models to identify the layout without relying on the raw view hierarchy~\cite{chen2020object, Zhang2021-yo}. For instance, \citet{Zhang2021-yo} proposed to use pixels to extract UI metadata such as UI element types and states. Similarly, others took a more classical computer vision approach~\cite{nguyen2015remaui, sun2020ui}, which detects a layout by identifying connected components from the pixels. However, to develop these layout prediction methods, in the first place, it is crucial to have UI datasets with clean layouts.

Compared to previous work, we focus on addressing the two major problems with view hierarchies, by detecting invalid objects and assigning each object a well-defined type. To this end, we create a clean layout dataset based on a public UI corpus, and develop a series of deep learning models to enable data cleaning in an automatic fashion.

\subsection{Mobile UI Screen Modeling}
Although our goal is to develop tools for improving dataset quality for future modeling tasks, our approach itself processes UI screenshots and raw view hierarchies and leverages UI modeling techniques. Here we briefly survey existing UI modeling techniques. 
\citet{zang2021multimodal} introduced a CenterNet-based model which, combined with text embeddings from the layout, predicts icons semantic types from screenshots. ~\citet{bai2021uibert} developed a pre-training model that facilitates multiple downstream tasks, including icon classification and app type prediction. Similarly, ~\citet{fu2021understanding} fed their own cleaned layouts to a Transformer model to perform a number of downstream tasks such as clickabilty prediction, relation prediction, or app classification. 
Finally, there is a rich body of work to connect natural language and mobile user interfaces. In Screen2Vec~\cite{li2021screen2vec}, Li et al. developed methods for embedding UI screens to enable tasks such as screen retrieval using nearest neighbors. 
\citet{li2020mapping} developed models that ground language instructions to executable actions on mobile phones. ~\citet{Li2020-la} developed models for generating captions for UI components on a mobile screen. Similarly, ~\citet{Wang2021-bh} proposed an approach for mobile UI screen summary, which describes the functionalities of the screen. \citet{Burns2021-ip} proposed a new task with a new dataset for automatic task completion based on mobile UI with iterative feedback.

There are several major deep architectures that have been used in these existing works. Computer vision models such as ResNet~\cite{he2015deep} are often used as the backbone for extracting features from images. Object detection models, such as Single-Shot multibox Detection (SSD)~\cite{Liu_2016}, Faster-RCNN~\cite{ren2015faster_rcnn}, or CenterNet~\cite{zhou2019objects} are often applied to detect UI objects on screenshots~\cite{Zhang2021-yo, chen2020object, zang2021multimodal}. Increasingly, Transformers~\cite{vaswani2017attention} have been used in a range of multimodal modeling tasks~\cite{li2020mapping, Li2020-la, Wang2021-bh}, which allows screenshot images and view hierarchy to be easily encoded via self-attention. 

Based on previous works, we design our models based on ResNet to encode images, and Transformer to perform cross-modal encoding and final decoding. We also investigate Graph Neural Networks~\cite{Gilmer2017} in this work as it can directly capture the tree structure of the view hierarchy.

\subsection{Data-Driven Automatic Cleaning Tools}
The need for developing data cleaning tools is ubiquitous~\cite{Ridzuan2019-rz}. There have been a number of previous efforts on developing automatic tools for data cleaning. For example, Chang et al.~\cite{Chang2017-ya} develop a tool for labeling datasets using crowd sourcing.
Cleanix~\cite{wang2014cleanix} is a tool that address abnormal value detection, incomplete data filling, deduplication, and conflict resolution in text-based data. SCAREd~\cite{yakout2013don} is an ML-based approach that attempts to learn correlations in correct text records, and predict adequate replacements in corrupted records.
In the same vein, KATARA~\cite{chu2015katara} aims at fixing inaccurate data by presenting a set of ML-issued corrections to crowd workers.
In the domain of vision and images, ~\citet{Ng2014-fa} propose a classifier to identify and remove outliers in a large scale face dataset. 
To the best of our knowledge, we are the first to propose automatic cleaning tools specifically targeted at mobile UI layout data.

\section{Dataset and type taxonomy}\label{sec:datasets}

To investigate our automatic approach for denoising layout data, we create a dataset of clean UI layouts based on an existing mobile UI corpus, dubbed as the CLAY dataset. In this section, we describe the dataset, the type taxonomy and the findings from the data collection.

\subsection{Mobile UI Corpus}
We use the open sourced Rico dataset~\cite{rico, Liu2018}, which contains 72K screenshots and view hierarchies from more than 9.7K different Android applications in 27 different app categories.

Each data point consists of a screenshot and layout information in view hierarchy about the objects on the screen. A view hierarchy is a tree structure where each node in the tree should correspond to an object on the UI. Each node contains a set of properties, such as the position of the UI object, its Android class, an optional content description, the resource identifier, and various attributes that characterize the object, e.g., whether the object is clickable or focusable.

\begin{figure*}
  \centering
  \includegraphics[width=1.\textwidth]{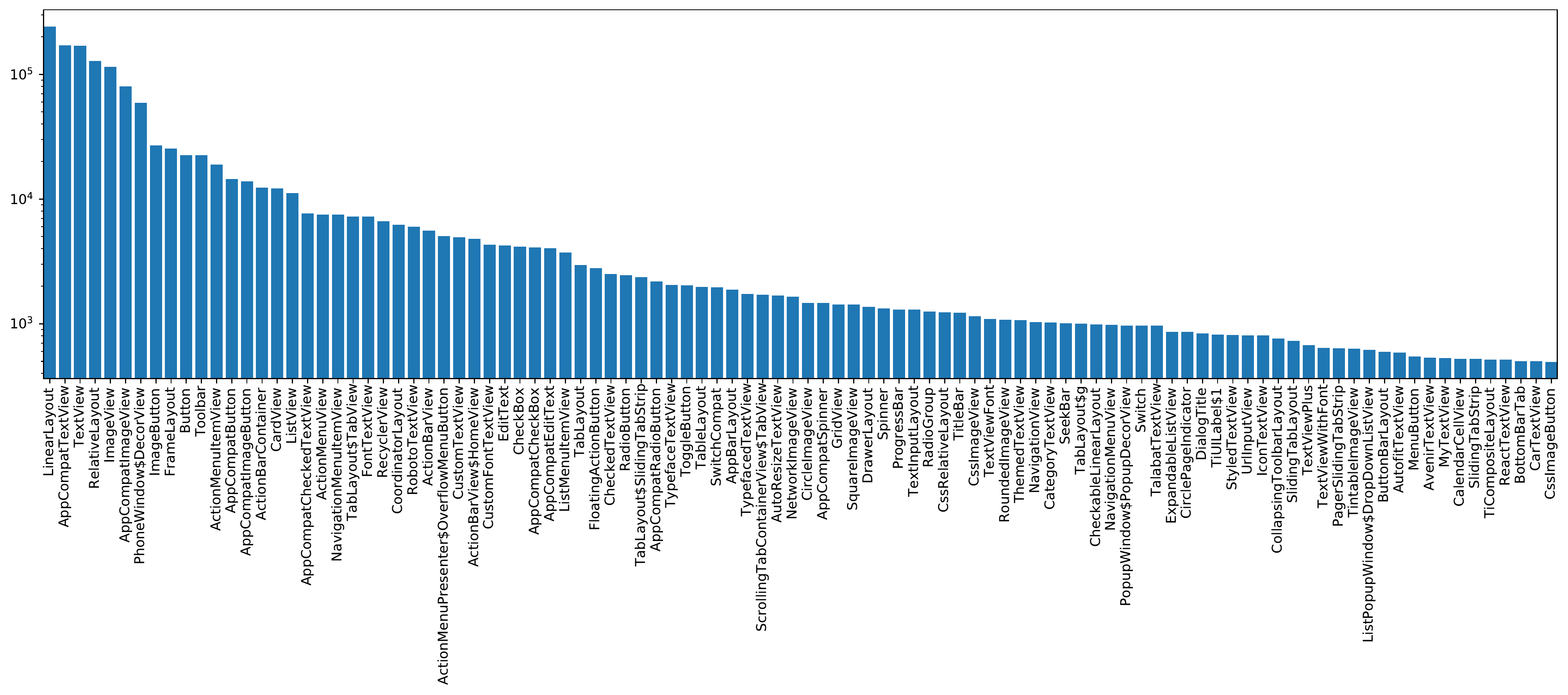}
  \caption{The log-scaled distribution of the 100 most popular Android classes in the Rico dataset. Generic types such as \texttt{LinearLayout}, \texttt{RelativeLayout}, or \texttt{FrameLayout} occur frequently in view hierarchies.}~\label{fig:android_class_hist}
  \Description{The distribution of top 100 Android classes in the Rico dataset. Many of them are generic types, e.g., LiearLayout, which don't provide much information of the UI element for models to learn.}
\end{figure*}

From the original Rico dataset, we removed layouts that contained no more than two objects as these layouts usually just contain one or two large container nodes and provide little information of the objects and structure on the screen. Before labeling, we preprocessed the view hierarchies to remove objects that are too narrow (width-height aspect ratio smaller than 0.01), too small (area smaller than 0.01\% of the screen), too large (area larger than the entire screen) or invisible based on the \textit{visible-to-user} and \textit{visibility} attributes. For objects with duplicate bounding boxes, we keep the one with a more specific type inferred from its Android class name or the last box in pre-order traversal as it is rendered at last. Occluded boxes are cut off to include only the visible part. Blank boxes with uniform color and empty containers are removed. We release the source code in the aforementioned GitHub repository so the results can be reproducible. This process resulted in a dataset with 59,555 screens.

Moreover, we counted 9,331 unique Android classes in the dataset. Many of the top classes are too generic to be useful information for object types. The 100 most popular Android classes are displayed in Figure~\ref{fig:android_class_hist}. There is a long tail distribution of app-specific object types, e.g., \texttt{ColombiaNativeAdView}, which convey little information of the object and are too sparse for model learning.

\subsection{Taxonomy and Labeling}
We define our type taxonomy based on the naming convention introduced previously in~\citet{Liu2018}, where \emph{semantic types} (e.g., \texttt{BUTTON}) are assigned to the UI components of the Rico view hierarchies to describe their functionalities. The previous taxonomy was defined based on an analysis of 720 screens.
Compared to the previous work~\cite{Liu2018}, we have introduced the changes described below. These changes are the result of multiple iterations on the original taxonomy. The rationale behind them was to provide classes that describe the visual appearance of the elements, and therefore we chose to merge elements that were visually similar.
For instance, we do not consider \texttt{VIDEO} and \texttt{IMAGE} as separate types as we assume them to be indistinguishable on static screenshots (this is true unless there is a visual cue such as a play button overlaid on the video).
Similarly, we have removed \texttt{WEB VIEW} and \texttt{MODAL}, and merged \texttt{MULTI-TAB} with \texttt{TOOLBAR}. 
We have also split \texttt{IMAGE} into two categories: \texttt{IMAGE}, which encompasses any natural image, photo or drawing; and \texttt{PICTOGRAM}, which represents an image containing vector graphics and a limited number of colors as found in icons and logos.

\begin{table*}
  \small
  \begin{tabular}{ll}
    \toprule
    Object Type & Description \\
    \midrule
    
    \texttt{ADVERTISEMENT} & Advertisement element in the screen \\
    \texttt{BUTTON} & A clickable element that allows the user to take actions and makes choices \\
    \texttt{CARD\_VIEW} & A container with a 'card' type of frame\\
    \texttt{CHECKBOX} & A square or circular shape that can be filled with a check or color. \\
    \texttt{CONTAINER} & A bounding box surrounding elements that belong to the same hierarchical group \\
    \texttt{DATE\_PICKER} & A calendar or date picker view\\
    \texttt{DRAWER} & An element that provides access to destinations and app functionality\\
    \texttt{IMAGE} & Natural image, photo or drawing \\
    \texttt{INVALID} & Invalid objects that don't have a valid visual representation, as described in Section 1 \\
    \texttt{LABEL} & A text element associated with an interactive element such as \texttt{TEXT\_INPUT}, \texttt{SWITCH}, \\
    & \texttt{CHECKBOX}, or \texttt{RADIO\_BUTTON}\\
    \texttt{LIST\_ITEM} & Repeated elements on the pages, with similar structure \\
    \texttt{MAP} & Map screens, either vector or satellite\\
    \texttt{NAVIGATION\_BAR} & An element that enables navigation through a series of hierarchical screens \\
    & (usually appears at the top of the screen). \\
    \texttt{NUMBER\_STEPPER} & A rolling wheel that allows to select preset values \\
    \texttt{PAGER\_INDICATOR} & A row of dots showing availability of multiple screens\\
    \texttt{PICTOGRAM} & An image containing vector graphics and limited number of colors (e.g., icons, logos, ...)\\
    \texttt{PROGRESS\_BAR} & A line or a circle that indicates a percentage of completion\\
    \texttt{RADIO\_BUTTON} & A circular shape that represents a single choice out of multiple options\\
    \texttt{SLIDER} & A control element that uses a knob or lever moved horizontally to control a variable\\
    \texttt{SPINNER} & An element with a drop down menu options\\
    \texttt{SWITCH} & An element with 2 toggle positions (usually on/off)\\
    \texttt{TEXT\_INPUT} & A text field where user can provide input\\
    \texttt{TEXT} & Text fields (a paragraph is considered a single text field)\\
    \texttt{TOOLBAR} & Element that contains buttons, icons, menus, and text \\
    & (usually appear at the bottom of the screen)\\

  \bottomrule
\end{tabular}
\caption{Our object type taxonomy and the description of each type.}~\label{tab:taxonomy}
\end{table*}

On the other hand, we also added a few elements such as \texttt{SPINNER} and \texttt{PROGRESS\_BAR}, as we considered them visually distinctive enough to justify new classes.
Finally, we added a more structural and hierarchical label with \texttt{CONTAINER}. 
A summary of our chosen taxonomy is shown in Table~\ref{tab:taxonomy}. We intend to label each valid node in the view hierarchy with one of these types.

\begin{figure*}
  \centering
  \includegraphics[width=1.\textwidth]{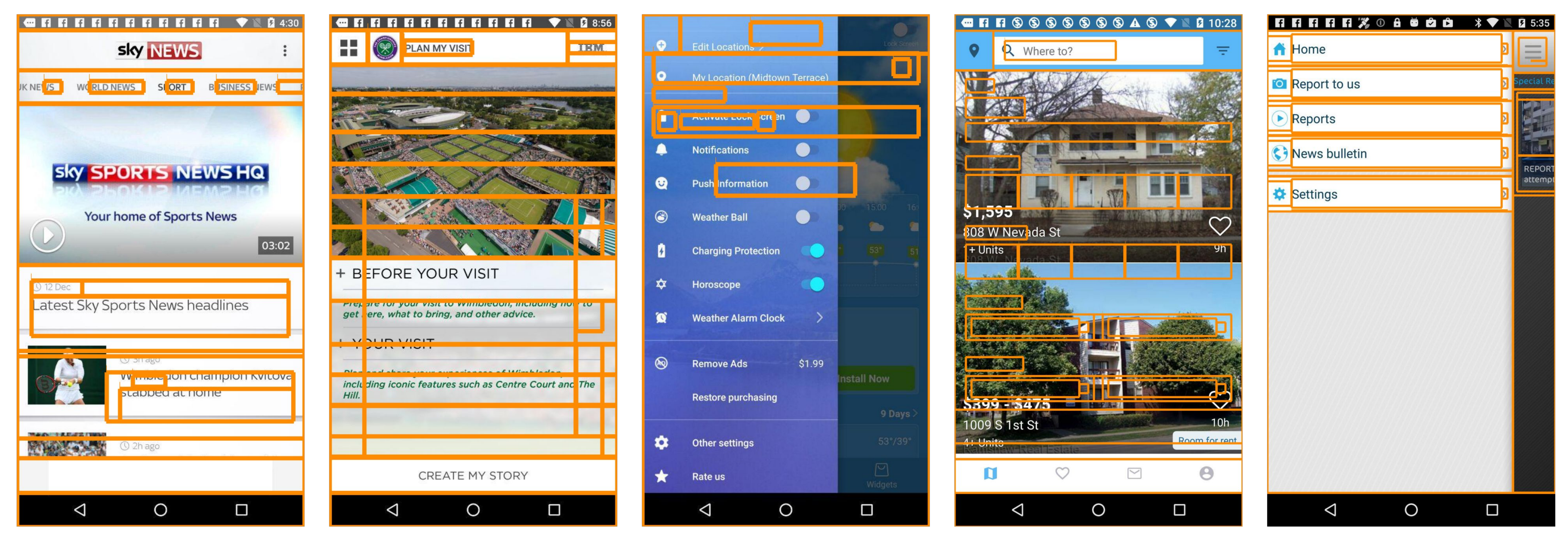}
  \caption{Examples of screens with visual mismatches between view hierarchy and the screenshots. The last screenshot shows elements in the background that cannot be interacted with; we consider these elements invalid too.}~\label{fig:layout_invalid}
  \Description{Examples of screens with invalid objects. These invalid objects have bounding boxes that misalign with rendered objects on the screens, as defined in Introduction.}
\end{figure*}

\begin{figure}
  \centering
  \includegraphics[width=0.5\textwidth]{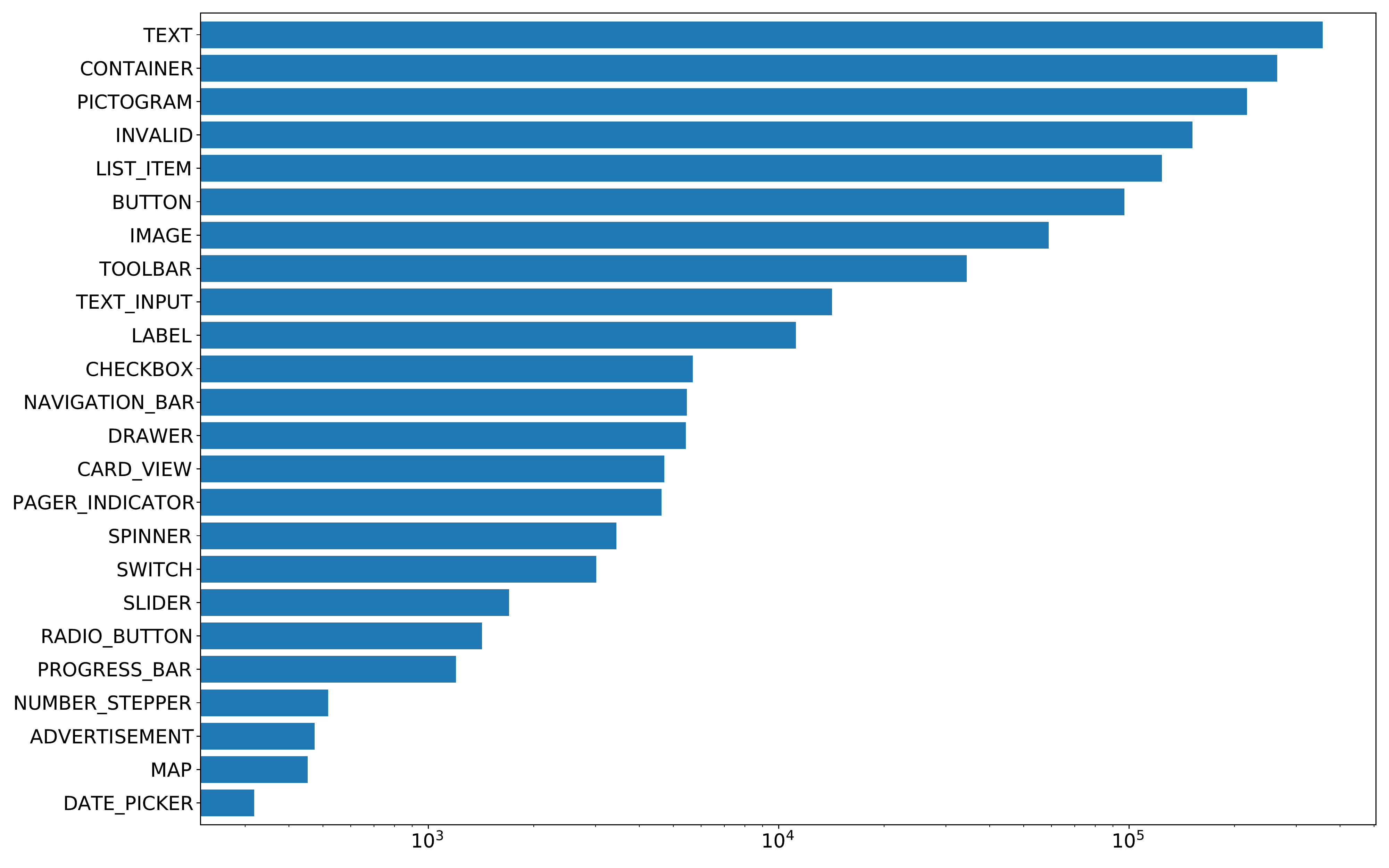}
  \caption{Object type distribution in the labeled CLAY dataset (log scale).}~\label{fig:label_dist}
  \Description{Labeled object type distribution. It shows our labels contain common object types, e.g., TEXT and CONTAINER, as well as rare types, e.g., MAP and DATE\_PICKER.}
\end{figure}

Based on the type taxonomy, we asked a group of 15 crowd human workers to label the filtered Rico dataset, which took at total of 1,577 hours.
We developed a web interface for human workers to annotate each view hierarchy element by assigning it with an appropriate type label. The interface shows a screenshot of the mobile interface, together with the bounding boxes extracted from the raw view hierarchy. Workers can choose the best type label from the list of the taxonomy, or flag the element invalid if its rendered bounding box does not correspond to a valid object on the screenshot. To ensure the quality, we audited the results by randomly sampling 3.1\% of the labeled examples during the labeling process and asked different labelers to verify them. It turned out 98.8\% of the audited objects were correctly labeled. Furthermore, for the validation and test set (see Table \ref{tab:dataset}), we labeled each object with 3 different labelers, and generated the final label by voting. After labeling the entire dataset that consists of 59,555 UI screens, 22,273 screens or 37.4\% of the dataset contain at least one invalid element (see~Figure~\ref{fig:layout_invalid}). The ratio of invalid versus valid objects is approximately 1:8. Figure \ref{fig:label_dist} shows the object type distribution of the labeled data. The dataset contains common UI objects including \texttt{TEXT} and \texttt{CONTAINER}, as well as rare types of objects, such as \texttt{DATE\_PICKER} and \texttt{KEYBOARD}.

\section{Task formulation and Model architecture}\label{sec:model_architecture}

We design a two-phase approach for denoising UI layout data. For the first phase, we propose a visual-based model, which detects invalid objects based on the object pixels.
For the second phase of object type recognition, we investigate two popular architectures: a GNN-based model~\cite{Gilmer2017} and a model based on the DeTR Transformer architecture~\cite{detr2020, deformable_detr2021}. Both of them are multi-modal, which rely on pixel information as well as raw view hierarchy to make predictions on the object type.


\subsection{Invalid Object Detection}

We first preprocess the layouts to filter out obviously invalid objects simply by looking at the layout tree and the rendering order of the objects. For example, objects fully occluded by other objects are removed, while the bounding boxes of objects partially occluded are trimmed.

\begin{figure}
  \centering
  \includegraphics[width=0.45\textwidth]{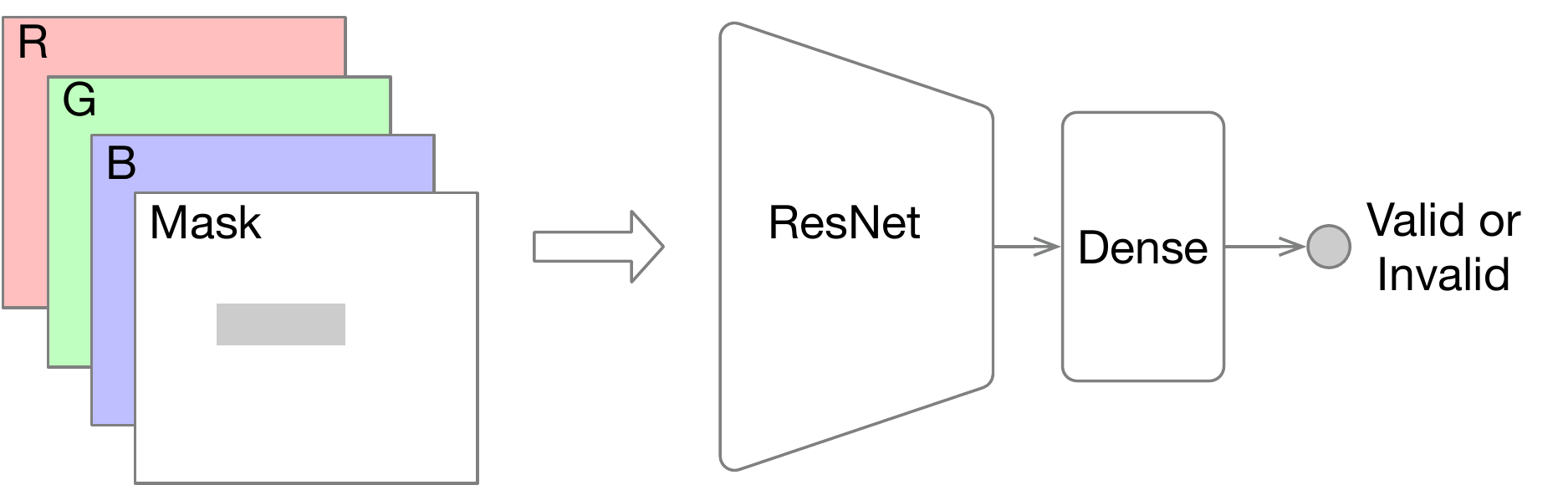}
  \caption{The binary classification model detects whether the object being inspected, as indicated by the mask channel, is invalid.}~\label{fig:binary_model}
  \Description{The binary classification model architecture. It shows how we mask the target UI element for prediction using the fourth channel in addition to RGB channels. A ResNet model is used to extract features for the binary prediction.}
\end{figure}

We then further filter out invalid objects using a binary classification model. We augment the popular ResNet model~\cite{he2015deep} with an extra input \emph{mask} channel, in addition to RGB, the three original image channels (see Figure~\ref{fig:binary_model}). 
The model examines one object in the layout at a time. With this extra mask channel, the input of the model is a matrix of size $[H, W, 4]$, with $H$ and $W$ as the height and width of the screenshot image. The first three channels correspond to the original pixels of the image, and the fourth channel indicates the bounding box of the object being inspected. The mask channel simply contains a binary mask with value $1$ at positions corresponding to the object bounding box, and $0$ otherwise. With the mask channel, the model is aware of the object location and focuses on the object pixels to make the prediction. In the meantime, the model has access to the context via the convolution operations in ResNet. The output of the model predicts how likely the object is invalid.

\subsection{Object Type Recognition}

In the second phase, we introduce two alternative deep learning approaches. We will discuss the pros and cons of each method in light of the experimental results. Both methods take the view hierarchy as input, and they use a similar approach for embedding each node (object) in the view hierarchy.

\subsubsection{View Hierarchy Node Embedding}

To represent each view hierarchy node as a dense vector, we embed its attributes separately and then combine these embeddings. For text-related information, we use the Android class name, \textit{content\_desc} and \textit{resource\_id} from each view hierarchy node. We use a vocabulary of size 28,536 to tokenize the text with the byte-pair encoding method, which is the same as BERT~\cite{Devlin2018-ia}. A maximum of first ten words of the three text fields are used for text embedding. These text embeddings are trained from scratch and max-pooled into a dense vector representing the information from the three text fields. At last, the text embedding of the node, denoted as $W$, is constructed by concatenating the three dense vectors.

To represent the object positional information, we use the four coordinates (i.e., the scalar values representing the left, right, top, and bottom location) of the object bounding box. Following \citet{Li2021-fourier}, each coordinate is mapped to a dense vector using fully connected layers and sinusoidal mapping. The four dense vectors are concatenated to form the positional encoding of the object, denoted as $P$. In this way the model can learn the representation of the coordinates via back-propagation for better performance.

\subsubsection{Type Recognition with Graph Neural Networks}

\begin{figure*}
  \centering
  \includegraphics[width=0.8\textwidth]{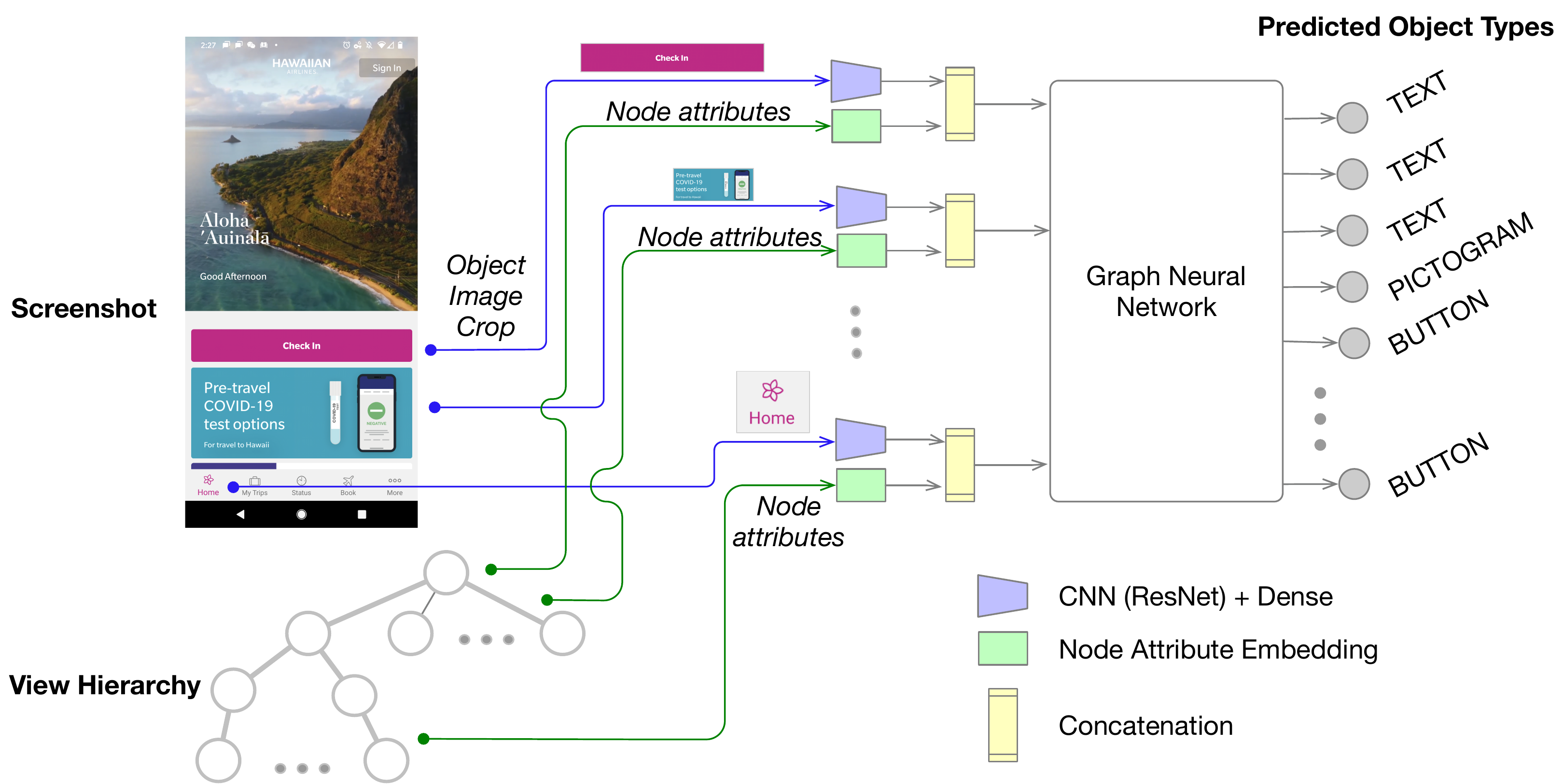}
  \caption{The architecture of our type predictor based on Graph Neural Networks (GNN). Each node is represented based on the concatenation of its node attribute embedding and the pixel encoding of its corresponding image crop. The CNN/Dense layers and the node attribute embedding layers are shared across all the nodes. These representations, along with the tree structure, are then processed by a multi-layer Graphical Neural Network that predicts the type of each node.}~\label{fig:layout_gnn}
  \Description{GNN model architecture for object type classification. Each layer of the GNN model will aggregate information for each node from its neighbors in the view hierarchy tree. At the end, each node will be represented by a dense vector containing information of itself and its neighbors for the final prediction.}
\end{figure*}

Our first proposed model for object type recognition is a multimodal GNN inspired by the message-passing neural network (MPNN) proposed by \citet{Gilmer2017}, which is a supervised-learning architecture that takes a graph as input. The output of the MPNN is a prediction of type for every node in the input graph. In our case, the input is the raw view hierarchy and the output is the object type of each node in the view hierarchy. The motivation for introducing a GNN-based approach is that view hierarchies are a tree structure, which is a special case of graphs, and thus GNNs can naturally leverage the structure.

To incorporate the pixel information of an object into the input, we crop the object pixels based on its bounding box in the view hierarchy as input to a ResNet-50 model. The output of the ResNet is flattened and passed through a dense layer to generate a dense vector, $I$, as the pixel encoding of the object.

In our MPNN, each node or object $o$ is represented by a hidden state $h_o^{t}$, where $t = 0, 1, \hdots,  T$ is the time step index. At $t=0$, the hidden state is initialized by concatenating the pixel, text and positional embeddings:

$$h_o^{0} = [I, W, P].$$

At every time step  $t = 1, 2, \hdots,  T$, a \emph{message kernel} $M$ is applied to every pair of connected objects, according to the view hierarchy tree structure.
For every node $o$, the messages from all its connections are gathered and aggregated via a pooling function, resulting in the vector $p_o^{t+1}$. The pooled vector is then fed to an $H$ kernel that updates the hidden state of a node at step $t+1$:

$$h_o^{t+1} = H(h_o^t, p_o^{t+1}).$$

Finally, a readout kernel $Y$ is applied to compute a logit for every node:

$$y_o = Y(h_o^T).$$

The variables of the kernels $H$, $M$, and $Y$, as well as the weights of the image encoder are all trainable parameters. An overview of our GNN model is depicted in Figure~\ref{fig:layout_gnn}. 

\subsubsection{Type Recognition with Transformer Models}

\begin{figure*}
  \centering
  \includegraphics[width=0.8\textwidth]{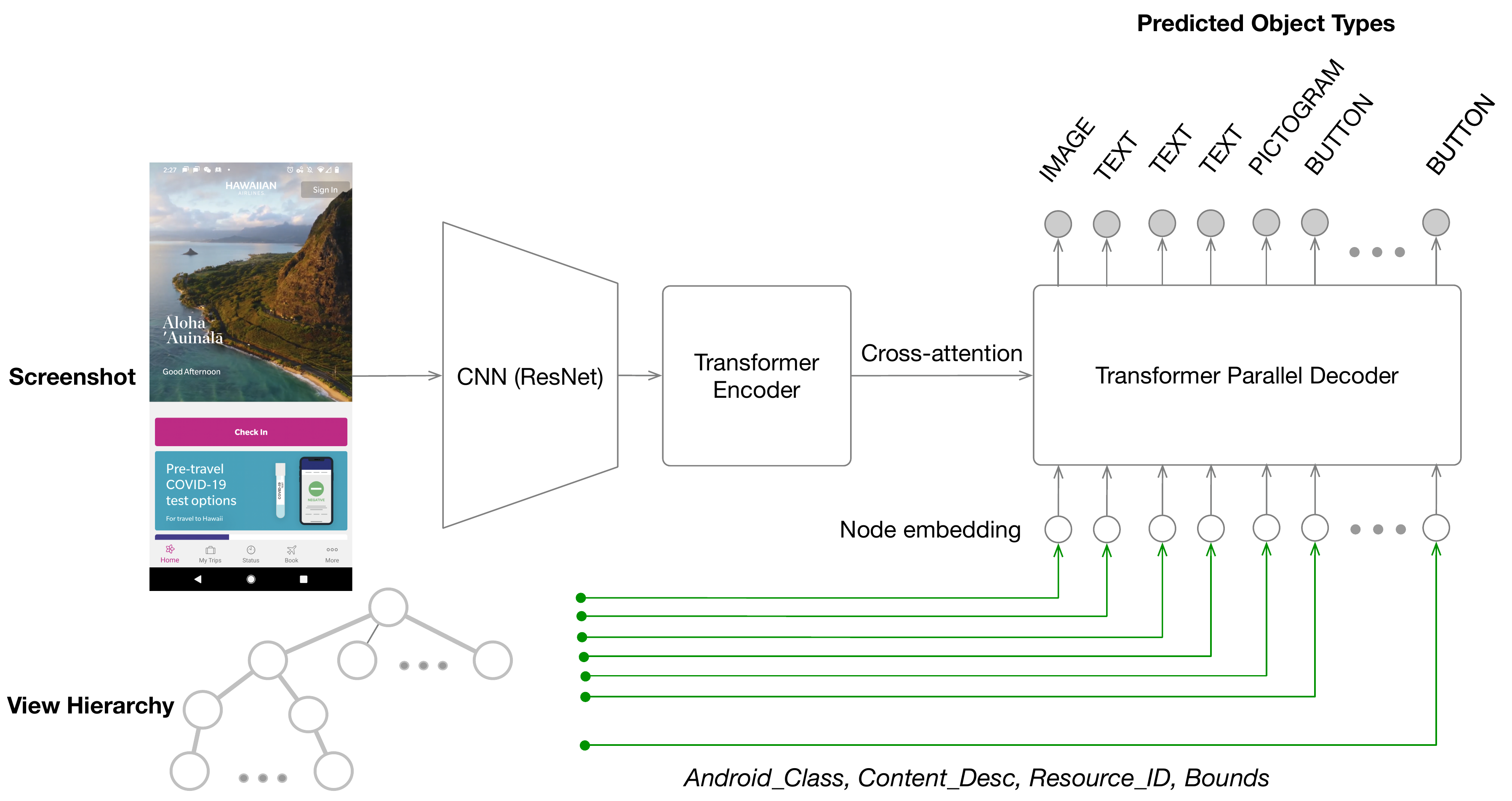}
  \caption{The architecture of the node type predictor based on the Transformer models. The entire screenshot is encoded by a CNN and Transformer encoder, and the Transformer Parallel Decoder then takes the view hierarchy nodes as input and predicts the type of each node.}~\label{fig:layout_detr}
  \Description{Transformer model architecture for object type classification. Each layer of the Transformer model will aggregate information for each node from all the nodes in the view hierarchy tree via self-attention. At the end, each node will be represented by a dense vector containing information of itself and other nodes for the final prediction.}
\end{figure*}

GNN directly captures the structure of a view hierarchy that induces a strong bias, which can be vulnerable to noisy structures. Thus, we design a Transformer-based model~\cite{vaswani2017attention}  that can learn object relationship via self-attention. In particular, we design our model based on DeTR~\cite{detr2020}, a model architecture that combines ResNet and Transformer for object detection. Instead of feeding in object queries as DeTR does, we feed in the view hierarchy node embedding as input to the parallel decoder. We also replace the output head that originally uses expensive Hungrian matching for object detection with the classification head for type prediction. At a high level, our model uses Transformer encoder stacked on ResNet to encode the entire screenshot image, and a Transformer parallel decoder to predict the object type of each node while attending to the image encoding. An overview of the model is shown in Figure \ref{fig:layout_detr}. 

The screenshot is first encoded with a ResNet-50 model, and the encoding is split into patches as inputs to the Transformer encoder. 
The outputs of the Transformer encoder is a matrix $M$ of shape $[N_m, H_m]$, where $N_m$ is the number of image patches and $H_m$ is the hidden state size of each patch. In the parallel decoder, each node or object $o$ is represented by a hidden state $h_o^{t}$, where $t = 0, 1, \hdots,  T$ is the layer index. The inputs to the parallel decoder is constructed by adding up the object text embedding and the positional encoding:

$$h_o^{0} = W + P.$$

\noindent The parallel decoder accesses the image encoding of the entire screenshot, $M$, by encoder-decoder attention. In each layer $t = 1, 2, \hdots,  T$, the parallel decoder will generate the hidden states for each object via self-attention of hidden states from last layer and cross encoder-decoder attention to the image:

$$h_o^{t+1} = \mathrm{cross\_attention}(\mathrm{self\_attention}(h_o^t), M).$$

\noindent Finally, a dense layer $Y$ is applied to compute the logits for every node:

$$y_o = Y(h_o^T).$$

\noindent For all our models of binary or multi-class classification, we train these models using the cross-entropy loss. $L2$ regularization is used for all the trainable weights in the model to mitigate over-fitting.

\section{Experiments}\label{sec:experiments}

In this section, we describe our experiments for evaluating the models and the results. We first experiment with the binary classification model for detecting invalid objects, and then evaluate the two multi-class classification models for object type recognition, in comparison with a baseline method that uses heuristics to predict UI object types.

\subsection{Dataset Splits}

We split our dataset of 59,555 screens randomly into the training, validation and test set. The split was performed package-wise, i.e., screens from the same package are not shared among the three splits. This is to avoid information leakage because screens from the same package might have similar layouts. Table \ref{tab:dataset} shows the statistics of the three sets.

\begin{table}[hp]
\small
\caption{Dataset statistics.}~\label{tab:dataset}
  \centering
  \begin{tabular}{l r r r}
    \hline
    {Split} & {Apps} & {Screens} & {UI Objects} \\
    \hline
   Training &  5,821 & 44,629 & 1,042,471 \\
   Validation & 989 & 6,207 & 139,411 \\
   Test & 1,698 & 8,719 & 186,501 \\
   \hline
   Total & 8,508 & 59,555 & 1,368,383 \\
    \hline
  \end{tabular}
\end{table}

\subsection{Model Configurations \& Training}
We here describe the configuration and training details of each model. Both the invalid object detection model and the GNN-based type recognition model are implemented in TensorFlow\footnote{\url{https://www.tensorflow.org}}. The Trans\-former-based model is implemented in JAX\footnote{\url{https://github.com/google/jax}}. We select the hyper-parameters to obtain the best performance on the validation set.

\subsubsection{Binary Classification Model for Invalid Object Detection} We train the model, based on ResNet-50, with a batch size of 1024 images for 15k steps to converge, with an initial learning rate 6e-4 and a reduced learning rate 6e-5 after 5.5k steps. To counter the skewed 8:1 distribution of valid and invalid objects, we re-sample the training data to have a ratio of 4:1 for valid and invalid objects, which has the best results on the validation set among the experiments using different ratios from 1:1 to 8:1. We do not apply resampling to the validation and test data. 

\subsubsection{GNN Models for Type Recognition}

We use a ResNet-50 model to encode the pixel information, the image crop of each UI object is resized to a squares of size $64 \times 64$, and the image embedding size is 32.
The GNN nodes are connected by bidirectional edges that represent the parent-child relationships of the elements in the layout. We also connect nodes that are spatially next to each other on the screenshot. At each step, 5 rounds of messages of size 32 are passed between the nodes. The messages are then aggregated with an attention pooling function.
The GNN model is trained to converge with 500K steps and a batch size of 32, with an initial learning rate of 2e-3, which is reduced to 1e-4 after 200K steps. 

\subsubsection{Transformer Models for Type Recognition} We use a ResNet-50 model as backbone, a 6-layer encoder for encoding the image, and a 6-layer parallel decoder to encode the view hierarchy objects and predict object types. The embedding dimension for view hierarchy objects is 256. For the Transformer encoder/decoder, we use 8 attention heads, MLP dimension 2048 and query/key/value dimension 256 \citep{vaswani2017attention}. The model is trained for 15k steps to converge, with a batch size of 128 examples, an initial learning rate 6e-5 for the ResNet backbone and 1e-4 for the Transformer encoder/decoder, and a reduced learning rate by 10 times after 5k steps. 

\subsubsection{Heuristic Baseline}\label{sec:heuristic_labeling}

As a baseline to compare our multiclass type recognition models, we implemented a heuristic method for inferring the layout types. Similar to the approach presented previously \citet{Liu2018}, the method deduces the object type from the Android class of the UI component in the view hierarchy. Since our type taxonomy is defined based on this previous work, we reused some of its mappings \cite{Liu2018}, and enhanced the inference rules based on the content description and resource id of the element. 
For instance, the \texttt{NAVIGATION\_BAR} type can be identified from the resource ids \texttt{android:id/navigationBarBackground} or \texttt{android:id/statusBarBackground}. Similarly, the \texttt{MAP} type can be detected from elements whose resource id is \texttt{com.google.and\-roid.apps.maps:id/map\_frame}. As we mentioned earlier, it is generally challenging to cover all the cases of mapping using a heuristic-based method. We release the code of the heuristic baseline for reproduction purposes in the aforementioned GitHub repository.

\subsection{Results}

We first report the model performance for the detection phase in Table \ref{tab:first_phase_acc}. Our model performs well for detecting the invalid objects, obtaining 82.7\% F-score with balanced precision and recall. This indicates that the visual-based model is effective for recognizing misaligned, invisible or grayed-out objects. The task is challenging because the ratio between invalid and valid objects is skewed. Balancing invalid and valid objects, i.e., positive and negative examples, in the training does significantly boost the model performance. We further analyze the quality of the model in Section~\ref{sec:error_analysis}.

\begin{table}[hp]
\small
  \centering
  \caption{The binary detection accuracy on test set.}~\label{tab:first_phase_acc}
  \begin{tabular}{l|c c c}
    \hline
    {Model} & {Precision} & {Recall} & {F-score} \\
    \hline
   ResNet &  83.3 & 82.0 & 82.7 \\

   \hline
  \end{tabular}
\end{table}

Next, we report the model performances for object type recognition in Table \ref{tab:second_phase_acc}. We average the scores across all the object types in two different ways: 1) weighted average where each type is weighted by the number of objects of that type, and 2) macro average where all the types have the same weight. Both GNN and Transformer-based model achieve significantly better performances than the heuristic baseline. The GNN model obtains better weighted average scores, which are dominated by the common object types. On the other hand, Transformer has better macro average scores. In Table \ref{tab:type_acc}, we can see that the GNN model has better performance for the more common types, while Transformer obtains more balanced scores across all types and performs better on rare types. Visual examples on various screenshots from the validation set are shown in Figure~\ref{fig:prediction_examples}.

\begin{table}
\small
\centering
\caption{The type recognition accuracy of each model on the test set.}~\label{tab:second_phase_acc}
\begin{tabularx}{0.47 \textwidth}{l|ccc|ccc}
    \hline
  & \multicolumn{3}{c|}{Weighted Average} & \multicolumn{3}{c}{Macro Average} \\
   \hline
    {Model}
    & {Precision}
    & {Recall}
    & {F1}
    & {Precision}
    & {Recall}
    & {F1} \\
    \hline
    Heuristic & 70.6 & 45.8 & 41.4 & 72.1 & 67.1 & 62.8 \\
    GNN & 86.1 & 85.9 & 85.9 & 83.6 & 74.8  & 78.3 \\
    Transformer & 85.1  & 84.6  & 84.7  & 84.2  & 79.5 & 81.4 \\
    \hline
\end{tabularx}
\end{table}

\begin{table*}
\small
\centering
\caption{The accuracy of predicting each object type for GNN and Transformer-based Model on the test set.}~\label{tab:type_acc}
\begin{tabularx}{0.58 \textwidth}{l|ccc|ccc}
  \hline
  & \multicolumn{3}{c|}{GNN} & \multicolumn{3}{c}{Transformer} \\
  \hline
    {Object Type}
    & {Precision}
    & {Recall}
    & {F-score}
    & {Precision}
    & {Recall}
    & {F-score} \\
    \hline

    ADVERTISEMENT & 52.6 & 21.7 & 30.8 & 85.2 & 50.0 & 63.0 \\
    BUTTON & 61.4 & 64.1 & 62.7 & 55.2 & 67.4 & 60.7 \\
    CARD\_VIEW & 87.4 & 64.1 & 73.9 & 73.2 & 72.9 & 73.0 \\
    CHECKBOX & 91.3 & 91.4 & 91.4 & 84.1 & 91.2 & 87.5 \\
    CONTAINER & 85.5 & 94.6 & 89.8 & 86.8 & 90.2 & 88.5 \\
    DATE\_PICKER & 92.0 & 71.9 & 80.7 & 100.0 & 81.2 & 89.7 \\
    DRAWER & 93.8 & 88.5 & 91.0 & 92.9 & 86.9 & 89.8 \\
    IMAGE & 72.9 & 64.8 & 68.6 & 75.6 & 69.8 & 72.6 \\
    LABEL & 78.3 & 75.0 & 76.6 & 60.0 & 55.6 & 57.8 \\
    LIST\_ITEM & 94.5 & 91.5 & 93.0 & 85.7 & 84.8 & 85.2 \\
    MAP & 78.4 & 47.1 & 58.8 & 82.3 & 65.9 & 73.2 \\
    NAVIGATION\_BAR & 86.6 & 88.6 & 87.6 & 84.0 & 88.4 & 86.2 \\
    NUMBER\_STEPPER & 94.1 & 81.4 & 87.3 & 87.9 & 86.4 & 87.2 \\
    PAGER\_INDICATOR & 85.3 & 55.1 & 67.0 & 78.1 & 67.2 & 72.3 \\
    PICTOGRAM & 85.1 & 85.1 & 85.1 & 87.3 & 85.5 & 86.4 \\
    PROGRESS\_BAR & 93.9 & 75.5 & 83.7 & 96.7 & 90.2 & 93.3 \\
    RADIO\_BUTTON & 54.0 & 67.9 & 60.2 & 79.4 & 82.7 & 81.0 \\
    SLIDER & 97.9 & 93.1 & 95.4 & 99.6 & 98.4 & 99.0 \\
    SPINNER & 79.5 & 55.4 & 65.3 & 86.9 & 62.0 & 72.4 \\
    SWITCH & 81.0 & 73.7 & 77.2 & 79.1 & 78.7 & 78.9 \\
    TEXT & 91.1 & 87.8 & 89.4 & 90.7 & 86.5 & 88.6 \\
    TEXT\_INPUT & 89.1 & 87.4 & 88.3 & 89.9 & 91.2 & 90.6 \\
    TOOLBAR & 97.3 & 95.9 & 96.6 & 96.6 & 96.2 & 96.4 \\
    
    \hline
\end{tabularx}
\end{table*}

\subsection{Error Analysis}\label{sec:error_analysis}

\begin{figure}[hp]
  \centering
  \begin{subfigure}[b]{0.49\columnwidth}
    \includegraphics[width=0.95\columnwidth]{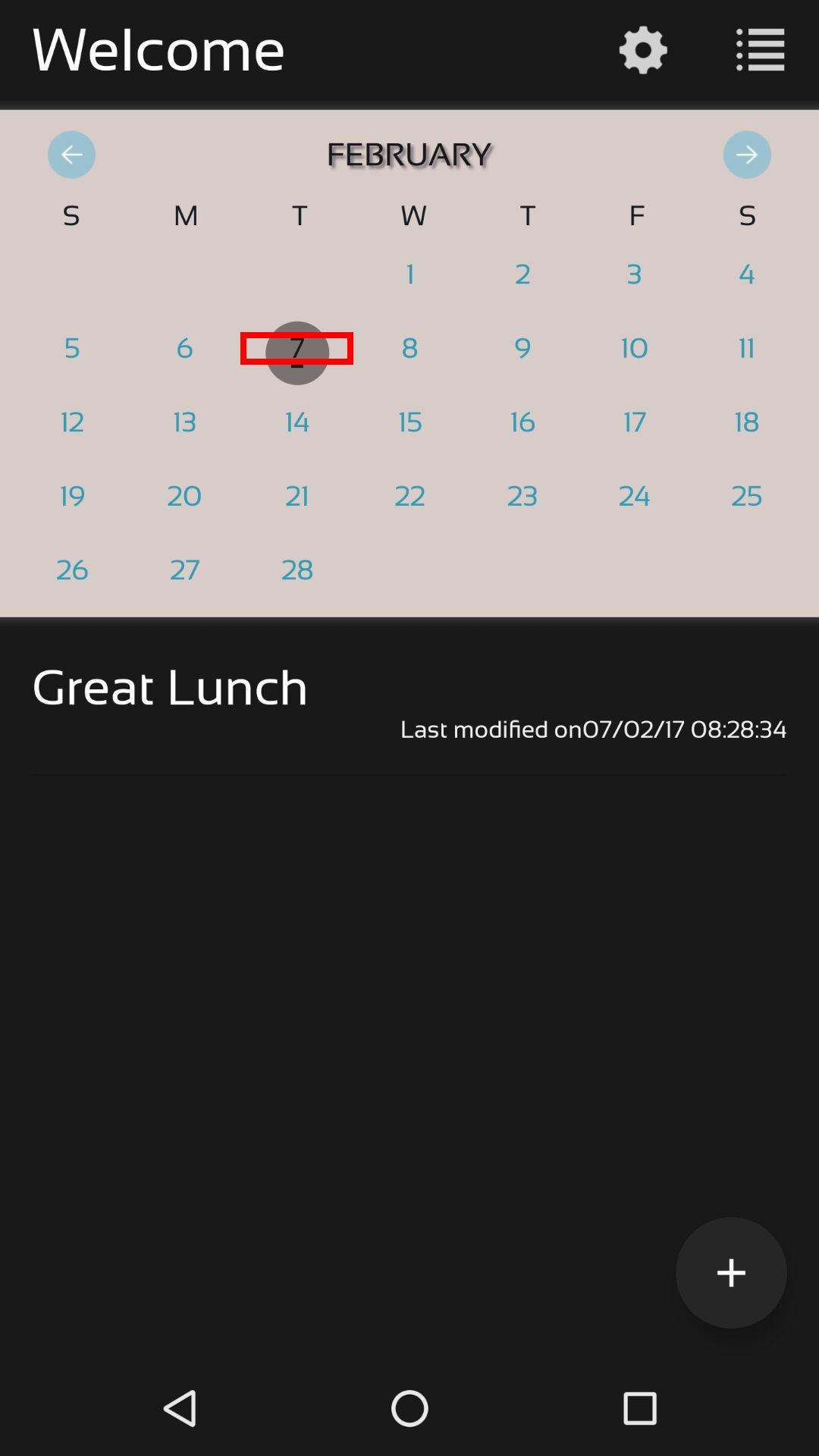}
  \end{subfigure}%
  \begin{subfigure}[b]{0.49\columnwidth}
    \includegraphics[width=0.95\columnwidth]{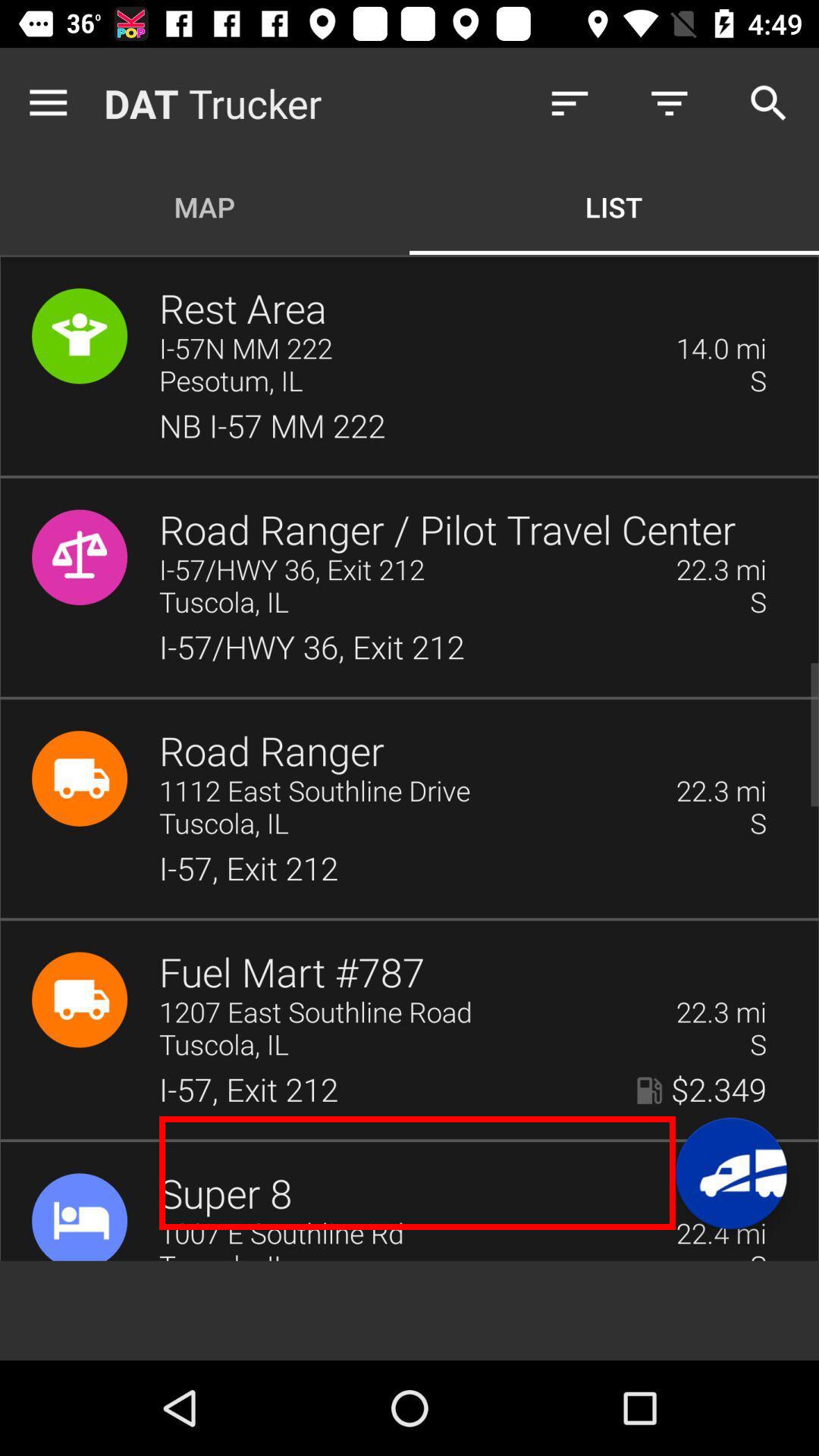}
  \end{subfigure}%
  \caption{False positive (left) and false negative (right) examples for the invalid object detection model.}~\label{fig:binary_error}
  \Description{Two error examples for the invalid object detection model. Both the false positive and the false negative example show that the model tends to make mistakes on ambiguous cases, which might be difficult even for human labelers.}
\end{figure}

We first analyze the errors of the invalid object detection model. We sample 100 (4.2\%) from all the false positive cases (valid objects predicted as invalid) and 100 (3.1\%) from all the false negative cases (invalid objects predicted as valid), and manually check the object on the screenshot to understand why the model makes the mistakes. Out of the 100 false positive errors, 47 cases have bounding boxes that are slightly shifted and partially cover the objects or do not cover the object tightly; 21 cases have bounding boxes overlapping with other objects; 14 cases are very small bounding boxes and the other 18 are due to reasons including blurred screen and confusion with background objects. The most common errors seem to be ambiguous cases which might be difficult even for a human labeler. For example, in the screenshot on the left of Figure \ref{fig:binary_error} the object with the red bounding box is labeled as valid by human workers, but predicted as invalid. In our labeling guideline, we define the invalid objects as those whose bounding boxes do not well align with the rendered objects, which leaves some uncertainty about how much misalignment is allowed for an object to be valid.

Similarly, out of the 100 false negative errors, 54 cases have bounding boxes that are shifted but encompass the object partially or not tightly. An example is shown on the right side of Figure \ref{fig:binary_error}, where the object in red bounding box is labeled as invalid by human workers but predicted as valid by the model. 22 cases are grayed-out objects in the background but the model failed to detect it; 17 cases have bounding boxes overlapping with other objects; 7 cases are very small bounding boxes.

\begin{figure*}
  \centering
  \includegraphics[width=0.65\textwidth]{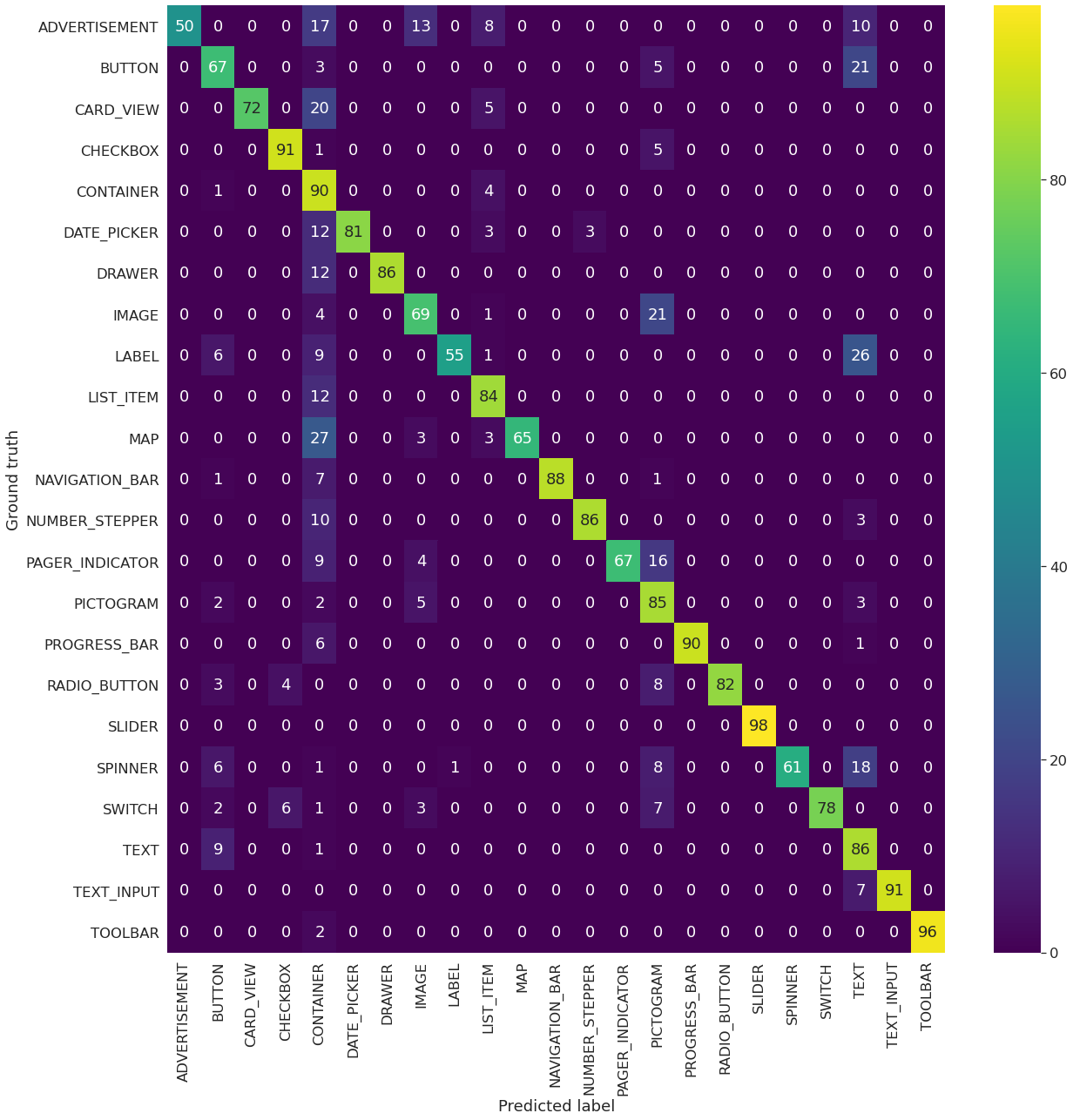}
  \caption{The confusion matrix of the object types for Transformer predictions on validation set. Each row is normalized by the number of groundtruth labels of the object types.}~\label{fig:confusion_matrix}
  \Description{The confusion matrix of the Transformer model results for object type prediction. The highlighted diagonal demonstrates that the model performs well for most types. We analyze the common confusion types in the Error Analysis subsection.}
\end{figure*}

\begin{figure*}
  \begin{subfigure}{0.19\textwidth} 
    \includegraphics[width=\linewidth]{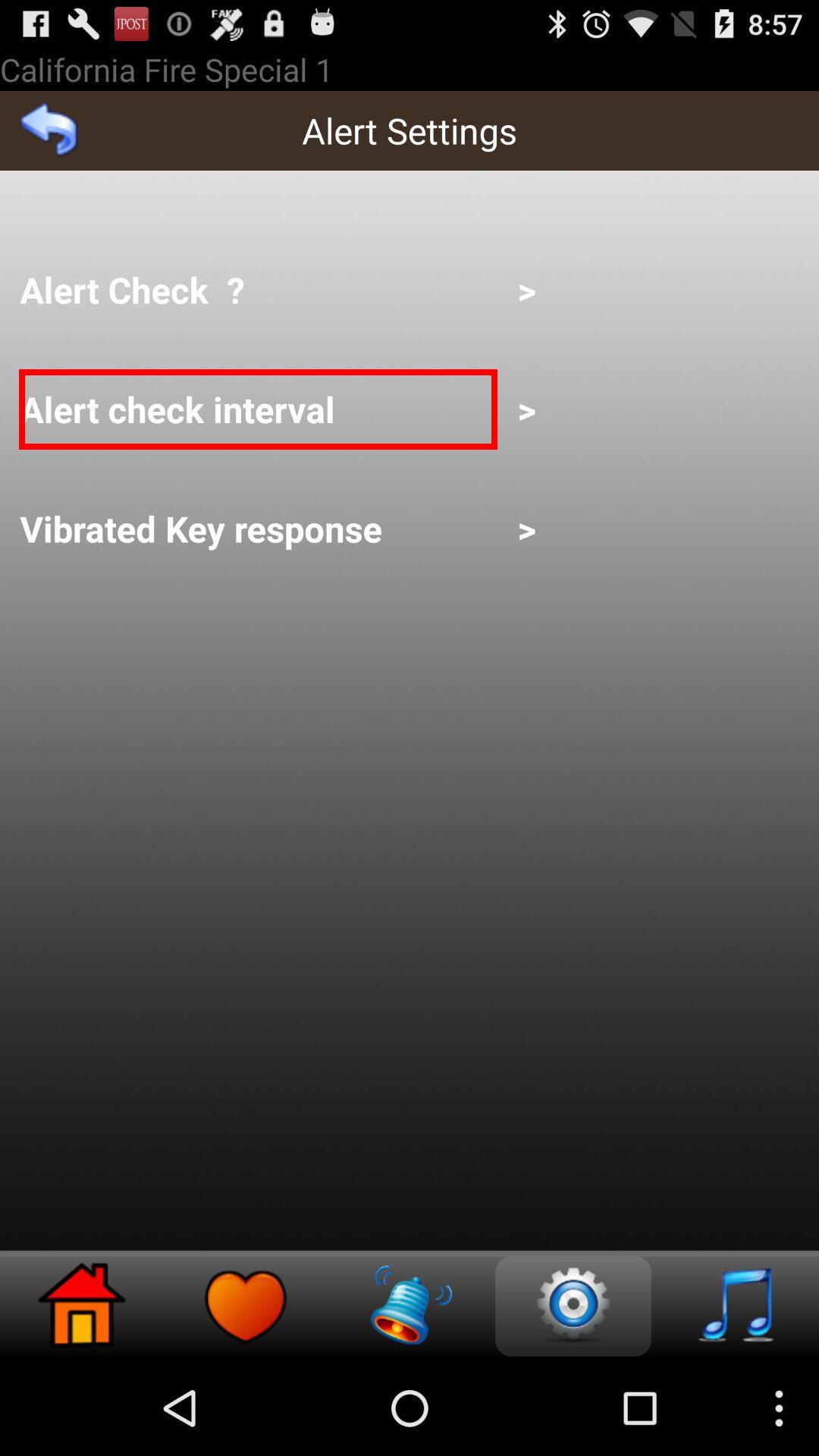}
    \caption{Ground truth: {\tiny BUTTON} \\ \hspace*{10.5pt}Prediction: {\tiny TEXT}} \label{fig:confusion_button_text}
  \end{subfigure}%
  \hspace*{\fill}   
  \begin{subfigure}{0.19\textwidth}
    \includegraphics[width=\linewidth]{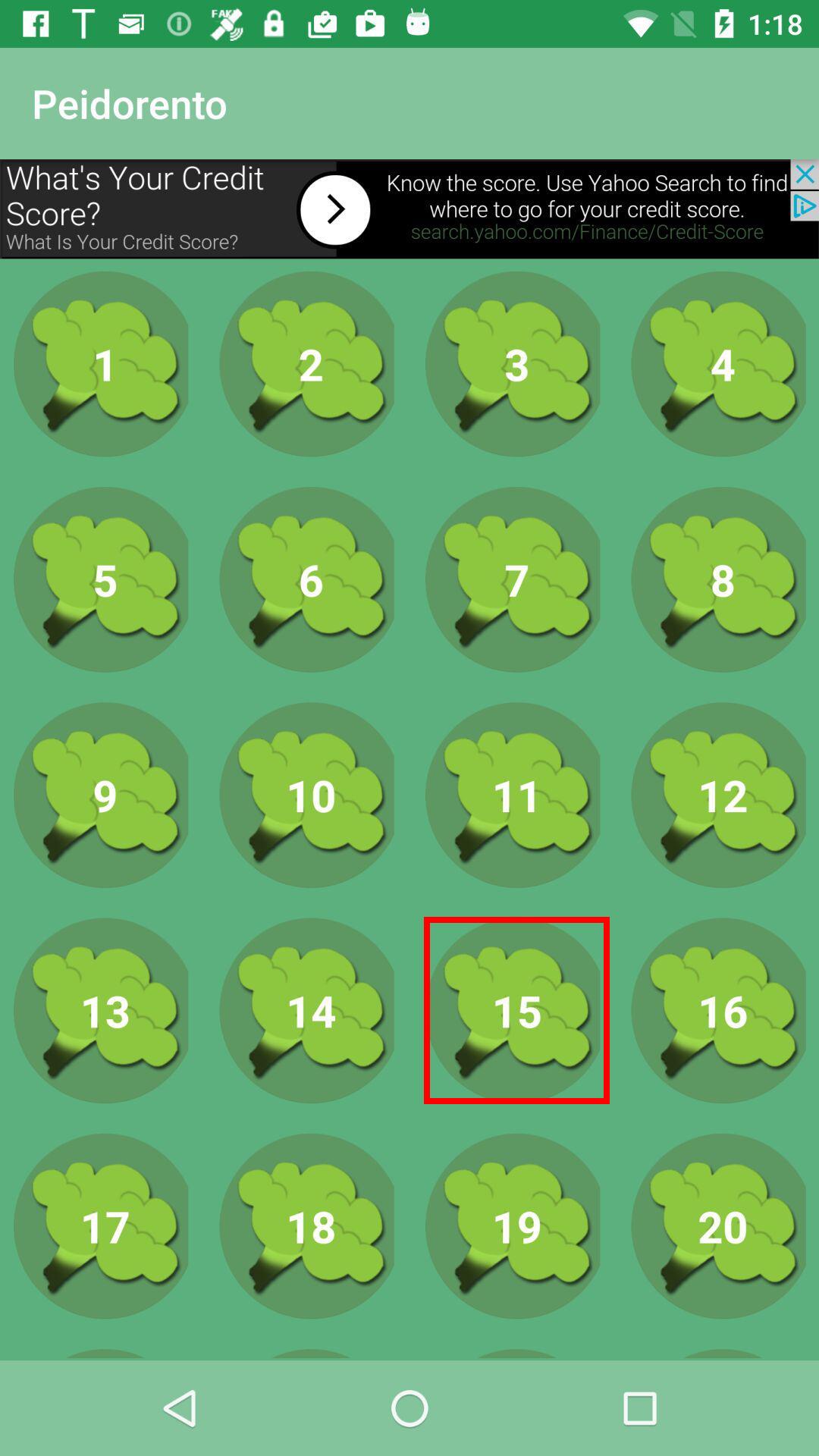}
    \caption{Ground truth: {\tiny IMAGE} \\ \hspace*{10.5pt}Prediction: {\tiny PICTOGRAM}} \label{fig:confusion_image_pictogram}
  \end{subfigure}%
  \hspace*{\fill}   
  \begin{subfigure}{0.19\textwidth}
    \includegraphics[width=\linewidth]{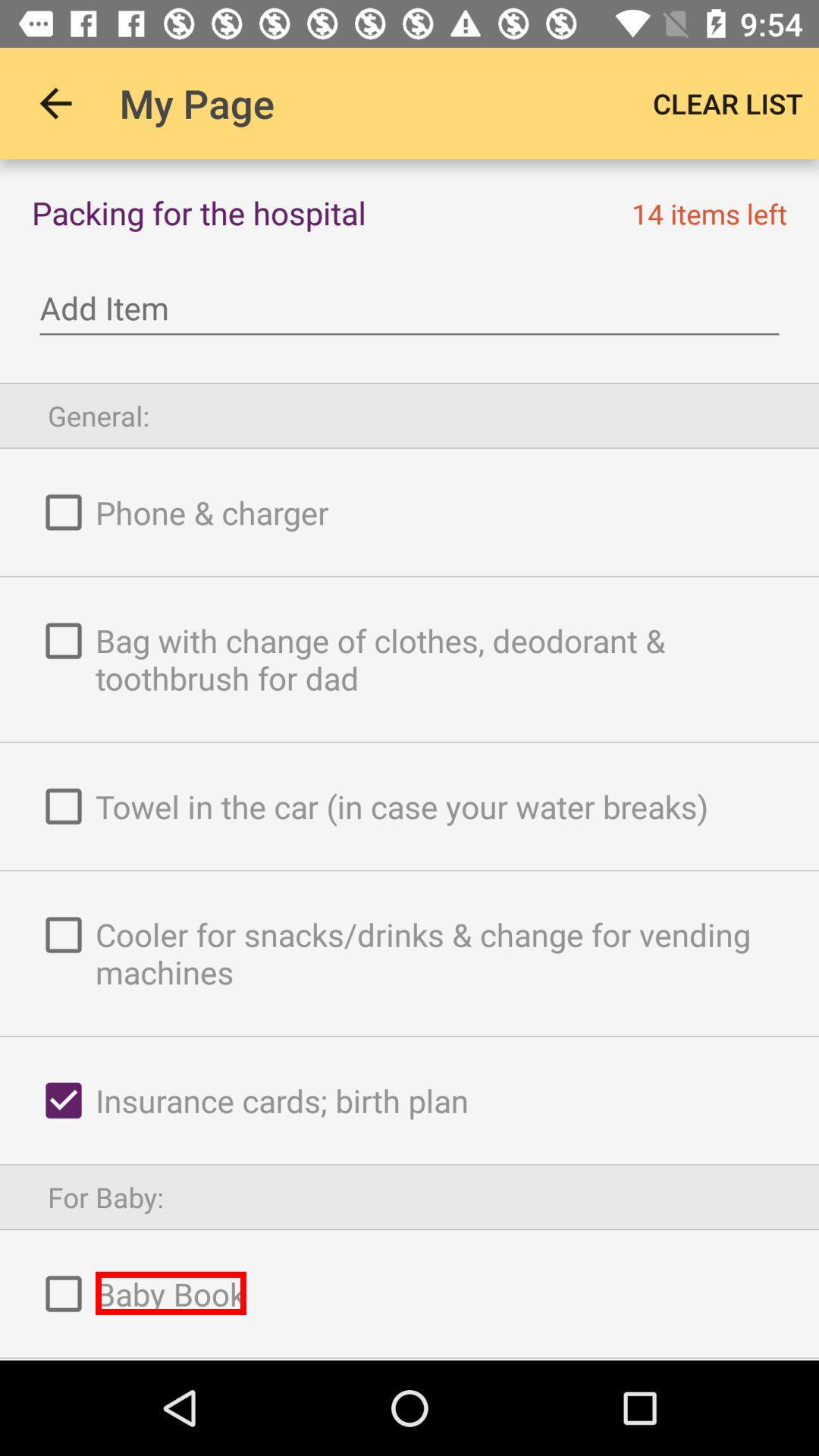}
    \caption{Ground truth: {\tiny LABEL} \\ \hspace*{10.5pt}Prediction: {\tiny TEXT}} \label{fig:confusion_label_text}
  \end{subfigure}
  \hspace*{\fill}   
  \begin{subfigure}{0.19\textwidth}
    \includegraphics[width=\linewidth]{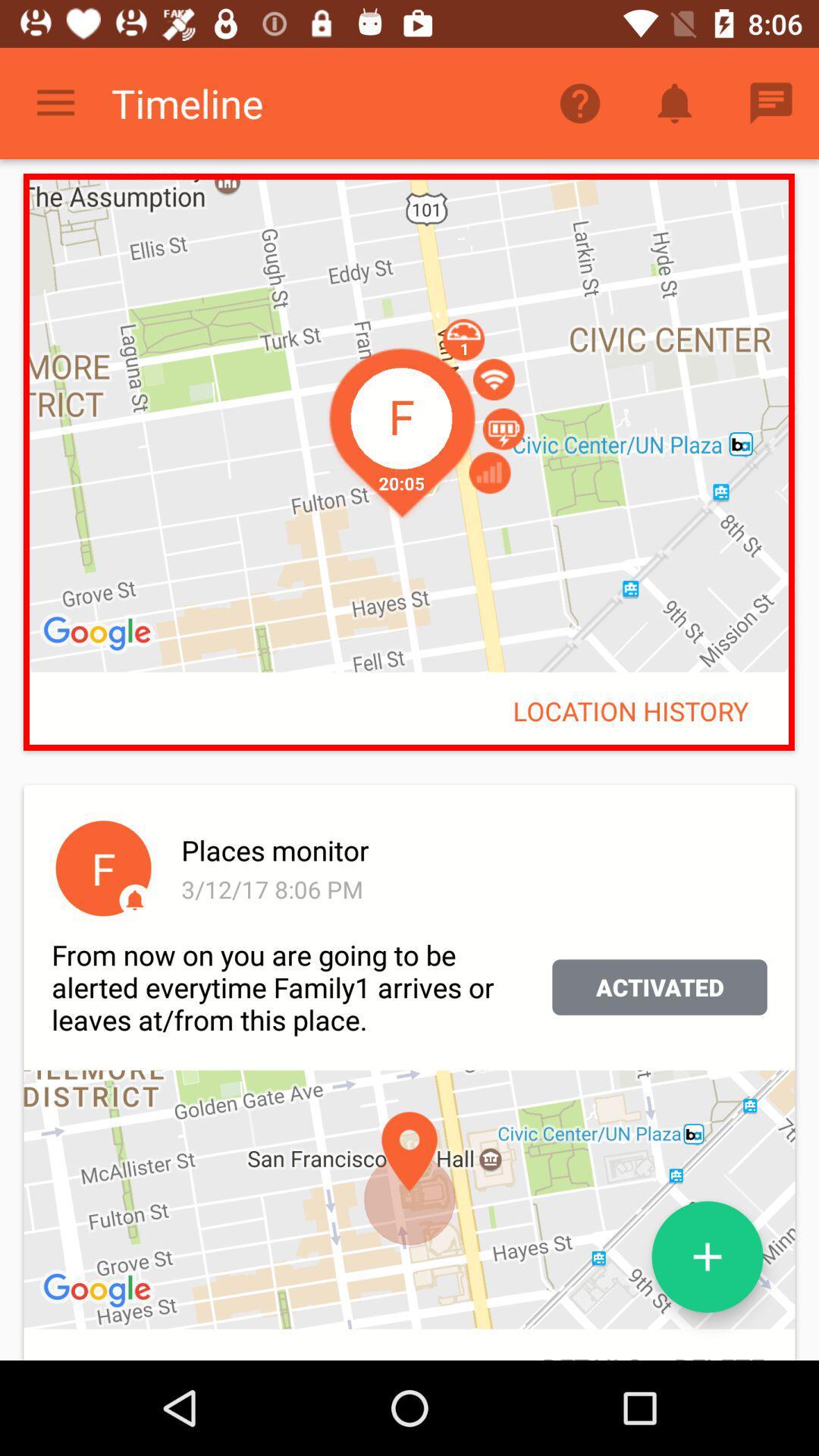}
    \caption{Ground truth: {\tiny MAP} \\ \hspace*{10.5pt}Prediction: {\tiny CONTAINER}} \label{fig:confusion_map_container}
  \end{subfigure}
\begin{subfigure}{0.19\textwidth}
    \includegraphics[width=\linewidth]{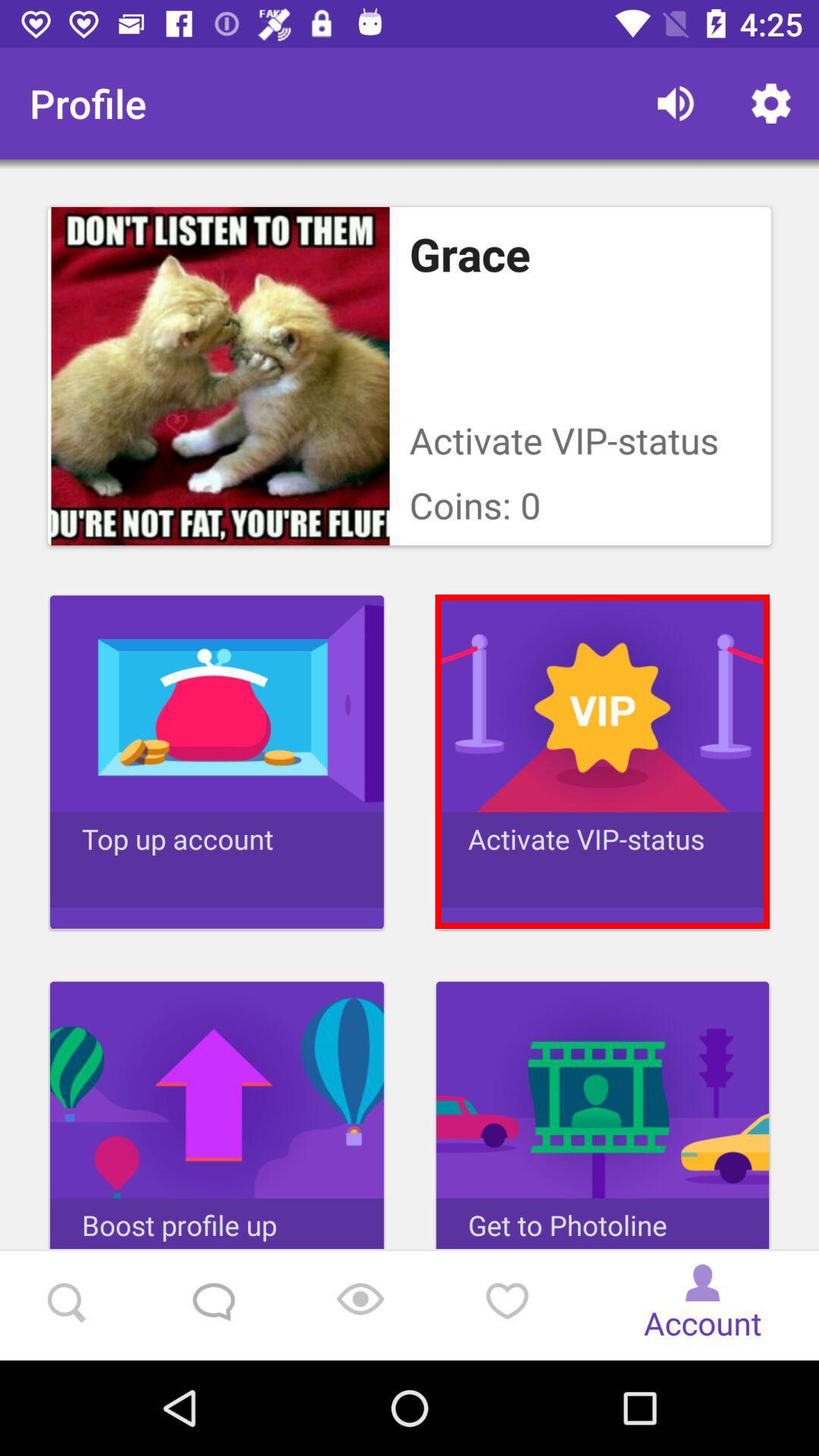}
    \caption{Ground truth: {\tiny CARD\_VIEW} \\ \hspace*{10.5pt}Prediction: {\tiny CONTAINER}} \label{fig:confusion_cardview_container}
  \end{subfigure}
\caption{Confusion examples for object type recognition.}~\label{fig:confusion_example}
\Description{Examples of the five most common confusion types. The confusions usually happen between object types that have similar appearance.}
\end{figure*}

For object type recognition, as shown in the confusion matrix (see Figure~\ref{fig:confusion_matrix}), our model performs well for most cases. The confusions tend to occur between several object types which can be ambiguous or similar looking. Among the 5 most common types of confusion, we examined all 27 instances of the confusion between \texttt{MAP} and \texttt{CONTAINER} and sampled 50 instances for the other 4 types, consisting of 1.9\% - 35.7\% of all the confusion instances. Figure~\ref{fig:confusion_example} illustrates the five types of confusions.
Specifically, \texttt{BUTTON} is confused with \texttt{TEXT} for some objects that look like pure text but actually can be clickable for users to take actions (Figure \ref{fig:confusion_button_text}). \texttt{IMAGE} is confused with \texttt{PICTOGRAM} on some icon-like images (Figure \ref{fig:confusion_image_pictogram}). \texttt{LABEL} is confused with \texttt{TEXT} for text-like \texttt{LABEL} objects (Figure \ref{fig:confusion_label_text}). The model predicts some instances of \texttt{CARD\_VIEW} and \texttt{MAP} as \texttt{CONTAINER}, possibly due to that \texttt{CARD\_VIEW} is a special type of container and might look similar (Figure \ref{fig:confusion_cardview_container}), and some large \texttt{MAP} objects contain other UI objects (Figure \ref{fig:confusion_map_container}). Data imbalance may be another reason as \texttt{CARD\_VIEW} and \texttt{MAP} have much fewer instances than \texttt{CONTAINER}, which makes it more difficult for the model to learn.

We further examined \texttt{ADVERTISEMENT}, for which both GNN and Transformer have lower scores. It is among the rare ones in the dataset, and usually has larger bounding boxes, for which the model confuses with \texttt{CONTAINER} or \texttt{IMAGE}. More training examples of such rare object types would potentially improve the model performance. We can merge some of the similar-looking object types to improve model performance when it is feasible for the downstream application.

\section{Discussions}\label{sec:discussions}

We created a large screen layout dataset based on 59,555 screens of Rico~\cite{rico}, with problematic objects flagged and more semantically meaningful types assigned to the valid objects. The cleaned layouts can be used for UI design research (e.g., similar layout retrieval \cite{Wang2021-bh}) and training new visual-based models \cite{sun2020ui} or data-cleaning models as described in this paper. The data-cleaning models can be used to preprocess the layout for downstream tasks, or clean large unlabeled UI dataset for training visual-based models, which can perform better with large-scale training data \cite{clip}.

Our models achieved an F1 score of 82.7\% for detecting invalid objects and an F1 score of 85.9\% for object type recognition. They offer a practical solution for cleaning datasets at scale. 
The invalid object detection model is effective despite the skewed ratio between valid and invalid objects, and made incorrect predictions for only a small amount of the objects when evaluated on the test set. Our proposed GNN and Transformer models perform comparably for weighted and macro average scores. For future work, combining the strengths of the two models to achieve good weighted and macro average scores is an interesting direction. We can try ensemble of the two models by joining the prediction probabilities. Encoding the structural information explicitly in the Transformer model might help the model to perform better on common objects. For the GNN model, accessing the entire image instead of the cropped pixels is another promising direction.


One limitation of our work is that we train and evaluate our models only on Android screenshots. This limitation is due to the lack of public corpus of screen layouts for these mobile platforms. The invalid object detection model relies on screenshot images only and might generalize better to other mobile OS, compared to the object typing models which rely on the OS-specific screen layouts.  We hope to include more diverse and recent screenshots in terms of packages and mobile OS in the future. 
For the denoising task, our model cleans up view hierarchy by labeling each node. However, more significant cleaning might be needed occasionally. For example, there might be objects that are rendered on the screen but missing in the view hierarchy. Therefore, adding new objects or adjusting objects' positions and attributes might be needed. Our DeTR-based model can include an object detection decoder \cite{detr2020} for this purpose, which would deserve further investigation.

\section{Conclusion}\label{sec:conclusion}

We present the CLAY pipeline, using a deep learning approach, for denoising mobile screen layouts, which are a critical data source for UI design research and UI semantic understanding. Our analysis reveals that automatically captured layouts in existing datasets are noisy and contain invalid objects and objects with noisy type information. To facilitate our investigation, we create the large CLAY dataset of clean UI layouts based on a public mobile UI corpus. We then propose a two-stage approach to first detect and remove invalid objects, and then classify the valid objects into the layout types defined in a taxonomy. Our experiments show that our models achieve good performance for both stages, which show a great potential to automatically denoise large layout datasets. These models will boost future efforts for large-scale UI modeling analysis. 



\bibliographystyle{ACM-Reference-Format}
\bibliography{references}

\appendix

\section{Prediction Examples}

\begin{figure*}
  \centering
  \includegraphics[width=0.85\textwidth]{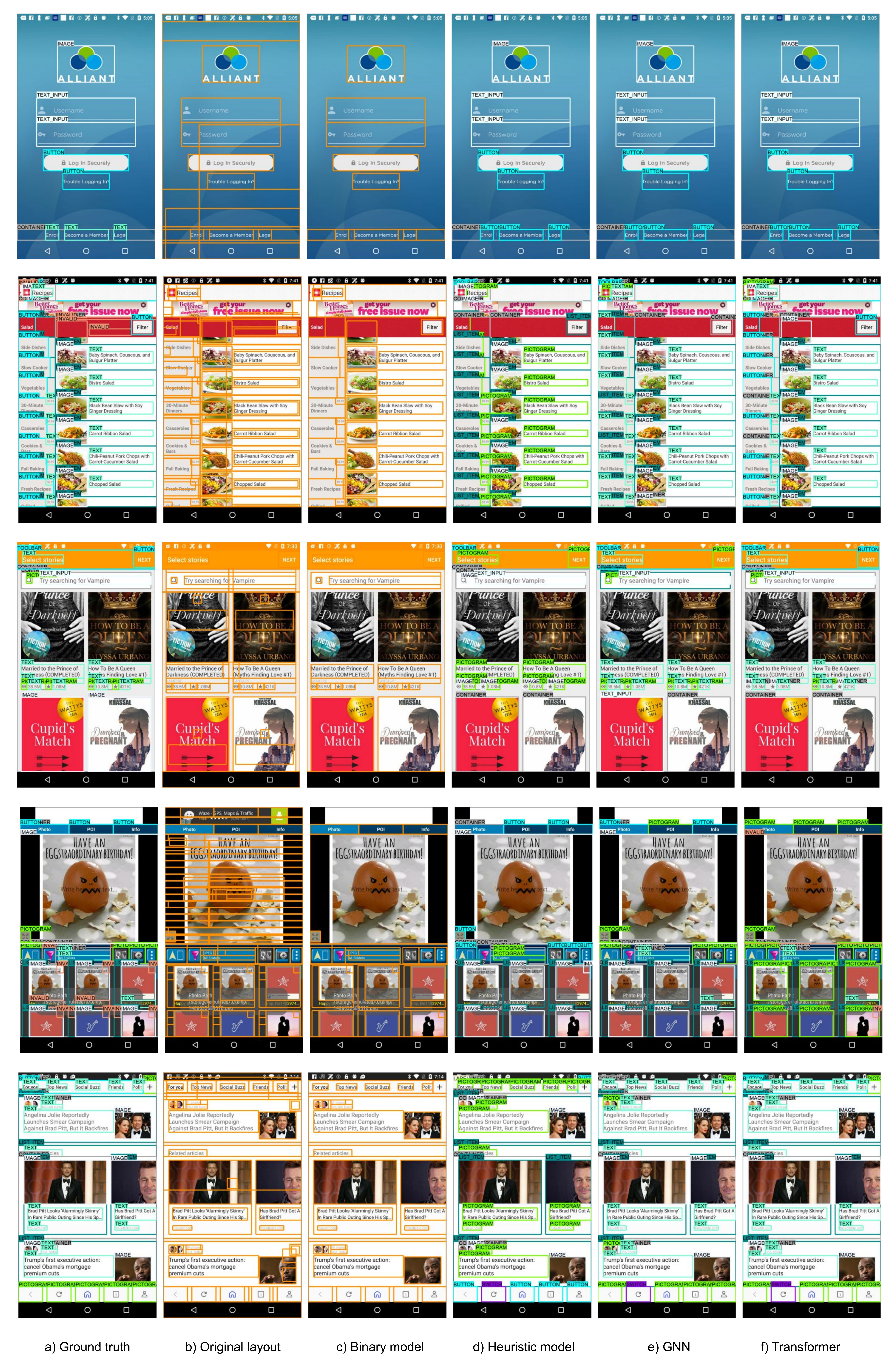}
  \caption{Visual examples of the predictions from our models on screenshots from the validation set. Column c shows outputs of the invalid detection model. Column d-f show outputs of the heuristic/GNN/Transformer model for object type recognition.}~\label{fig:prediction_examples}
  \Description{Prediction examples of our models. The comparison between the original layout and the model outputs demonstrate that our models are effective in removing invalid objects and classifying semantic object types, resulting in much cleaner and meaningful UI layouts.}
\end{figure*}

Figure \ref{fig:prediction_examples} shows five example screens with the original layouts and the outputs from our models.

\end{document}